\documentclass[12pt,a4paper]{article}
\usepackage{amsmath}
\usepackage{amssymb}
\usepackage{amsxtra}
\begin{document}
\newtheorem{theorem}{Theorem}
\author{Ilja Schmelzer\thanks
       {WIAS Berlin}}

\title{Derivation of the Einstein Equivalence Principle in a Class of
Condensed Matter Theories} \sloppypar

\maketitle

\begin{abstract}

We consider a class of condensed matter theories in a Newtonian
framework with a Lagrange formalism related in a natural way with the
classical conservation laws

\begin{eqnarray*}
\partial_t \rho + \partial_i (\rho v^i) &= &0 \\
\partial_t (\rho v^j) + \partial_i(\rho v^i v^j + p^{ij}) &= &0
\end{eqnarray*}

We show that for an algebraically defined ``effective Lorentz metric''
$g_{\mu\nu}$ and ``effective matter fields'' $\varphi$ these theories
are equivalent to material models of a metric theory of gravity with
Lagrangian

\[
L = L_{GR} + L_{matter}
  - (8\pi G)^{-1}(\Upsilon g^{00}-\Xi (g^{11}+g^{22}+g^{33}))\sqrt{-g}
\]

which fulfils the Einstein equivalence principle and leads to the
Einstein equations in the limit $\Xi,\Upsilon\to 0$.

\end{abstract}

\section{Introduction}

General relativity is not only a very successful theory of gravity,
but also a very beautiful one.  Part of its beauty is the remarkable
fact that there are many different ways to general relativity.  On one
hand, there are remarkable formulations in other variables (ADM
formalism \cite{ADM}, tetrad, triad, and Ashtekar variables
\cite{Ashtekar}) where the Lorentz metric $g_{\mu\nu}$ appears as
derived. Some of them may be considered as different interpretations
of general relativity (like ``geometrodynamics'').  On the other hand
we have theories where the Lorentz metric is only an effective metric
and the Einstein equations appear as some limit (spin two field in QFT
on a standard Minkowski background \cite{Feynman} \cite{Weinberg65}
\cite{Deser}, Sakharov's approach \cite{Sakharov}, string theory
\cite{Polchinski}).

An effective curved Lorentz metric appears also in classical and
quantum condensed matter theory, as the ``acoustic'' metric ``seen''
by phonons.  This has been shown, for example, for an irrotational
flow of a barotropic fluid in \cite{VisserAcoustic}.  Following Unruh
\cite{UnruhAcoustic} such acoustic metrics have been considered as toy
models to study Hawking and Unruh radiation (see \cite{Rosu} and
references there).  Very interesting is also the consideration of
superfluid $^3He-A$, which shows not only effective gravity, but also
effective gauge and chiral fermion fields, where relativistic and
gauge symmetry appears in a certain approximation \cite{Volovik}
\cite{Volovik98}.  In this context, the question arises how far the
analogy between the ``acoustic metric'' and general relativity goes.
At a first look, it seems that it cannot be taken very far.  Visser
\cite{VisserAcoustic} writes that ``the aspect of general relativity
which does not carry over to the acoustic model is the dynamics -- the
Einstein equations''.  Volovik \cite{Volovik98} mentions that the
equations are ``highly contaminated by many non-covariant terms''.

The results of this paper suggest that a much stronger analogy may be
possible.  We introduce here a quite general class of
condensed-matter-like theories.  We require classical conservation
laws (continuity equation, Euler equation), Lagrange formalism, and a
special relation between them in agreement with Noether's theorem.
This is already sufficient to prove the Einstein equivalence principle
for ``effective matter fields'' on an ``effective metric''.  We derive
a general Lagrangian, which differs from GR in only two additional
terms which depend on the Newtonian background frame.  As a
consequence, in a certain limit we obtain the classical Einstein
equations.

The derivation is based on a slightly unorthodox but beautiful variant
of covariant description of a theory with preferred frame -- we use
the preferred coordinates $X^\alpha(x)$ as variables.  This gives
Euler-Lagrange equations for the preferred coordinates $X^\alpha(x)$.
In case of translational symmetry they become conservation laws.  This
observation motivates our main physical assumption: the identification
of the Euler-Lagrange equation for $X^\alpha$ with the classical
conservation laws of condensed matter theory.  We introduce an
effective metric $g_{\mu\nu}$ with an algebraic definition similar to
ADM decomposition.  This transforms the conservation laws into the
harmonic condition.  We obtain a remarkable composition of three
coordinate-related features: ADM decomposition, harmonic coordinates,
and conservation laws.  Based on this transformation, the derivation
of the EEP and the Lagrangian is surprisingly simple.

Then we discuss the derivation itself and its possible applications.
The most surprising thing in this derivation is our set of
assumptions.  They look quite innocent and natural for a condensed
matter theory.  Simplicity and beauty of the derivation are remarkable
too.

This connection between condensed matter theory and theory of gravity
may be used in two directions.  First, we may try to apply it in
condensed matter theory.  While our assumptions seem to be natural for
condensed matter theories, they require a Lagrange formalism in
unusual form.  This suggests that the connection may be used to find
new Lagrange formalisms for condensed matter theories starting with GR
Lagrangians.

The other direction would be the consideration of condensed matter
theory in a Newtonian framework as a candidate for a more fundamental
theory behind general relativity.  Essentially this would be a revival
of old, pre-relativistic ether theory.  The ability to derive the
Einstein equivalence principle and the beauty of this derivation
removes some important arguments against this concept.  And there are
several independent domains where a Newtonian framework gives
advantages: We obtain well-defined local energy and momentum
conservation laws.  There is no ``problem of time'' which makes
quantization much easier.  A preferred frame is also a necessary
prerequisite for realistic causal hidden variable theories of quantum
theory like Bohmian mechanics \cite{Bohm} or Nelson's stochastics
\cite{Nelson}.  Therefore, a revival of the old ether concept may be
not as unreasonable as it sounds. 

But, independent of the viability of ether theories of such type, the
relation between the equations of classical condensed matter theory
and the Einstein equivalence principle is a remarkable and beautiful
observation which is worth to be considered in more detail.

\section{Covariant Management of Preferred Coordinates}

Let's start with the description of our variant of covariant
description of theories with a Newtonian framework.  We have to be
careful here to avoid confusion about the meaning of ``covariance''.
It is well-known today that ``general covariance'' is not a special
property of general relativity, as initially thought by Einstein.
Instead, every physical theory allows a formulation which does not
depend on the choice of coordinates \cite{Kretschmann}.  The
difference between general relativity and other theories is the
absence of an a priori, absolute geometry \cite{Anderson}.  An example
of a covariant formulation of Newtonian mechanics can be found, for
example, in \cite{MTW}.

We consider here some class of condensed matter theories in a
classical Newtonian framework, thus, theories with an an absolute, a
priori geometry.  We have an absolute preferred foliation defined by
absolute time $T = X^0$, and we have an absolute Euclidean background
metric which is $\delta_{ij}$ in our preferred coordinates $X^i$.

To describe this absolute background in a covariant way different
variables may be used.  One possibility is to handle the constant,
flat metric of the background geometry like a variable metric tensor
$\gamma_{\mu\nu}(x)$ and to add the covariant equation
$R^\mu_{\nu\kappa\lambda}[\gamma]=0$.  Now, our choice of variables is
different -- we use the preferred coordinates $X^\alpha$ themself as
the variables which describe the Newtonian background.  Last not
least, the preferred coordinates are functions $X^\alpha(x)$ on the
manifold and may be used as other scalar functions to describe a
physical theory.

An important point is that we require that the $X^\alpha$ should be
used in the Lagrangian and the Lagrange formalism {\em as usual scalar
fields\/}.  A short but sloppy formulation would be that the
Lagrangian depends on the preferred coordinates in a covariant way.
Unfortunately this may cause confusion.  More accurate, the Lagrangian
should depend on the functions $X^\alpha(x)$ like a usual covariant
scalar Lagrangian depends on scalar fields $U^\alpha(x)$.  To minimize
confusion, let's name such a Lagrangian ``weak covariant''.  Instead,
a ``strong covariant'' Lagrangian in this formalism is also ``weak
covariant'', but, moreover, does not depend on the preferred
coordinates $X^\alpha(x)$ at all -- which would be covariant in the
usual sense of general relativity.

What it means can be best seen in a simple example.  Assume we have a
scalar product in the preferred coordinates defined by
$(u,v)=\delta_{ij}u^i v^j$.  This expression is non-covariant.  The
same scalar product can be rewritten as
$(u,v)=\delta_{ij}X^i_{,\mu}X^j_{,\nu}u^\mu v^\nu$.  This expression
is already weak covariant.  Note that the preferred coordinates
$X^\alpha$ themself are scalar functions.  Their upper index $\alpha$
is not a spatial index, but simply enumerates the four scalar
functions.  Thus, $u^0$ is non-covariant because of the open upper
spatial index 0, but $X^0_{,\mu}u^\mu$ is already weak covariant.

Despite the special geometric nature of coordinates (see appendix
\ref{appCoordinates}) we can apply the Lagrange formalism as usual and
obtain Euler-Lagrange equations for the preferred coordinates
$X^\alpha$.  Now, these equations have a close connection with the
conservation laws for energy and momentum.  We know from Noether's
theorem that symmetries of the Lagrangian lead to conservation laws.
Especially, conservation of energy and momentum follows from
translational symmetry in the preferred coordinates.  Now, for
translational symmetry $X^\alpha\to X^\alpha+c^\alpha$ we do not need
any theorem to find them -- we obtain them automatically.  The
Lagrangian does not depend on the $X^\alpha$ themself, only on their
partial derivatives.  Therefore the Euler-Lagrange equations for
$X^\alpha$ {\em are\/} already conservation laws:

\begin{theorem}\label{Noether1}

If the Lagrangian has translational symmetry $X^\alpha \to X^\alpha +
c^\alpha$, then the Euler-Lagrange equations for the preferred
coordinates $X^\alpha$ are conservation laws:

\[ \frac{\delta S}{\delta X^\alpha} = \partial_\mu {T^\mu_\alpha} = 0 \]

\end{theorem}

We know that in agreement with Noether's second theorem these
conservation laws disappear if we have general covariance, that means,
in our formalism, strong covariance.  Again, in our formulation we do
not need any theorem to see this -- if there is no dependence on the
$X^\alpha$, the Euler-Lagrange equation for the $X^\alpha$, which are
the conservation laws, disappear automatically:

\begin{theorem}\label{Noether2}

If L is strong covariant, then

\[ \frac{\delta S}{\delta X^\alpha} \equiv 0 \]

\end{theorem}

In above cases, the proof is so obvious that there seems no need to
write it down.  Of course, there is not much to wonder about, we have
simply used a set of variables $X^\alpha$ appropriate for the symmetry
groups we have considered -- translational symmetry.  Which is, of
course, the reason why we have introduced this formalism here.  Above
symmetry groups as well as the conservation laws play an important
role in the following.

In the reverse direction, the last theorem does not hold.  There are
Lagrangians which are not strong covariant but have covariant
Euler-Lagrange equations, like the well-known Rosen Lagrangian
\cite{Rosen}.  This is a consequence of the fact that the Lagrangian
is not uniquely defined by its Euler-Lagrange equations.  But the
Rosen Lagrangian is a well-known GR Lagrangian too.  So, what defines
the most general GR Lagrangian
\footnote{Here, ``most general'' means that also higher order
derivatives of the metric and non-minimal interactions are allowed.
This is in agreement with the modern concept of effective field
theory, as, for example, expressed by Weinberg \cite{Weinberg}: ``I
don't see any reason why anyone today would take Einstein's general
theory of relativity seriously as the foundation of a quantum theory
of gravitation, if by Einstein's theory is meant the theory with a
Lagrangian density given by just the term $\sqrt{g}R/16\pi G$.  It
seems to me there's no reason in the world to suppose that the
Lagrangian does not contain all the higher terms with more factors of
the curvature and/or more derivatives, all of which are suppressed by
inverse powers of the Planck mass, and of course don't show up at
energy far below the Planck mass, much less in astronomy or particle
physics.  Why would anyone suppose that these higher terms are
absent?''} in our formalism is not strong covariance, but the weaker
property

\[ \frac{\delta S}{\delta X^\alpha} \equiv 0. \]

In the following we assume that the Lagrangian is given in (or can be
transformed into) weak covariant form.  In appendix \ref{appGeneral}
we consider the question if this is a non-trivial restriction or not.
Nonetheless, it seems worth to note that until now we have only
introduced a general formalism which is in no way restricted to the
condensed matter application below.

\section{Description of the Condensed Matter Theories}

The class of theories we consider here describe something similar to a
classical medium in a Newtonian framework -- Euclidean space and
absolute time.  This ``medium'' is described by steps of freedom
typical for condensed matter theory: a positive density $\rho(x,t)$, a
velocity $v^i(x,t)$, and a symmetrical tensor field $p^{ij}(x,t)$
similar to pressure, but negative definite.  As usual in condensed
matter theory, there may be also various material properties
$\varphi^m(x,t)$.  These material properties are not specified or
restricted in our class of theories.

We require the existence of a weak covariant Lagrange formalism.  Now,
on one hand from theorem \ref{Noether1} follows that in case of
translational symmetry the Euler-Lagrange equations for $X^\alpha$ are
conservation laws.  On the other hand, we know the classical
conservation laws from condensed matter theory -- the continuity
equation and Euler equation.  It seems quite natural and innocuous to
identify them.  Thus, we identify the continuity equation with the
equation for time $T = X^0$:

\begin{equation} \label{continuity}
\frac{\delta S}{\delta X^0} \sim
 \partial_t \rho + \partial_i (\rho v^i) = 0
\end{equation}

and the Euler equation with the equations for spatial coordinates
$X^i$:

\begin{equation} \label{Euler}
\frac{\delta S}{\delta X^j} \sim
 \partial_t (\rho v^j) + \partial_i(\rho v^i v^j+p^{ij})=0
\end{equation}

The other equations (the ``material equations'') are not specified.
{\em That's already all.\/}

Nonetheless, it should be noted that in this step we have made
important physical assumptions: we have only a single, universal
medium, and no interactions, especially no momentum exchange with
other media or external forces.

\section{The Effective Metric and Effective Matter}

The new variable we introduce is the ``effective metric'' $g_{\mu\nu}$.
It is defined algebraically by the following formulas:

\begin{eqnarray*}\label{gdef}
 \hat{g}^{00} = g^{00} \sqrt{-g} &=  &\rho \\
 \hat{g}^{i0} = g^{i0} \sqrt{-g} &=  &\rho v^i \\
 \hat{g}^{ij} = g^{ij} \sqrt{-g} &=  &\rho v^i v^j + p^{ij}
\end{eqnarray*}

Note that this decomposition of $g^{\mu\nu}$ into $\rho$, $v^i$ and
$p^{ij}$ is a variant of the ADM decomposition \cite{ADM}.  The
signature of $g^{\mu\nu}$ follows from $\rho>0$ and negative
definiteness of $p^{ij}$.  It is a global hyperbolic metric --- the
Newtonian background time is a global hyperbolic function.

Now, a remarkable observation which is essential for our derivation is
what happens with our four classical conservation laws in these new
variables.  They simply (essentially by construction) become the
harmonic condition for the effective metric $g_{\mu\nu}$:

\[ \partial_\mu (g^{\alpha\mu}\sqrt{-g}) = \square X^\alpha = 0 \]

That means, the preferred coordinates of the Newtonian background are
the harmonic coordinates of the effective metric $g_{\mu\nu}$.

All other (unspecified) material properties $\varphi^m(x)$ of the
condensed matter theories are interpreted as ``effective matter
fields''.  That means, different condensed matter theories with
different sets of material properties (inner steps of freedom) lead to
different effective matter fields.  Instead of the original condensed
matter Lagrangian $L=L(\rho,v^i,p^{ij},\varphi^m)$, we have now a
Lagrangian $L=L(g^{\mu\nu},\varphi^m)$ which depends on an effective
metric $g_{\mu\nu}$ and effective matter fields $\varphi^m(x)$.

\section{Derivation of the General Lagrangian}

Now, it's time to derive the general form of the Lagrangian for our
class of condensed matter theories in these effective variables.
Let's at first introduce some notations for the proportionality
factors used in the relations between the Euler-Lagrange equations for
the $X^\alpha$ and the continuity and Euler equations.  We need two
constants $\Xi, \Upsilon$, and we introduce them for convenience with
a common factor $(4\pi G)^{-1}$:

\begin{eqnarray}
\frac{\delta S}{\delta T}   &=& - (4\pi G)^{-1} \Upsilon \square T\\
\frac{\delta S}{\delta X^i} &=&   (4\pi G)^{-1} \Xi \square X^i
\end{eqnarray}

Introducing a diagonal matrix $\gamma_{\alpha\beta}$ by
$\gamma_{00}=\Upsilon,\gamma_{ii}=-\Xi$ we can write these equations
in closed form as
\footnote{Note that the coefficients $\gamma_{\alpha\beta}$ are only
constants of the Lagrange density, the indices enumerate the scalar
fields $X^\alpha$.  They do not define any fundamental, predefined
object of the theory.  Instead, variables of the Lagrange formalism,
by construction, are only $g_{\mu\nu}$, $\varphi^m$ and $X^\alpha$.}

\[\frac{\delta S}{\delta X^\alpha}
	\equiv-(4\pi G)^{-1}\gamma_{\alpha\beta}\square X^\beta\]

Now, we can easily find a particular Lagrangian which fulfills this
condition:

\[L_{0} = -(8\pi G)^{-1}\gamma_{\alpha\beta}X^\alpha_{,\mu}X^\beta_{,\nu}
       	g^{\mu\nu}\sqrt{-g}\]

Then, let's consider the difference  $L-L_{0}$.  We obtain:

\[ \frac{\delta \int(L-L_{0})}{\delta X^\alpha} \equiv 0 \]

But this is simply the condition that the Euler-Lagrange equations of
the Lagrangian $L-L_0$ are strong covariant, that means, the equations
of general relativity.  Thus, $L-L_0$ is simply the most general
Lagrangian of general relativity.  So we obtain:

\[L = -(8\pi G)^{-1}\gamma_{\alpha\beta}X^\alpha_{,\mu}X^\beta_{,\nu}
       	g^{\mu\nu}\sqrt{-g}
    + L_{GR}(g_{\mu\nu}) + L_{matter}(g_{\mu\nu},\varphi^m).
\]

This Lagrangian fulfils the Einstein equivalence principle (EEP,
cf. \cite{Will}) in its full beauty -- we have a metric theory of
gravity.  Having derived the Lagrangian, we have {\em derived\/} this
principle for the whole class of condensed matter theories which
fulfill our assumption.

In the preferred coordinates $X^\alpha$ (where we have
$X^\alpha_{,\mu}=\delta^\alpha_\mu$) we obtain a non-covariant form of
the Lagrangian:

\[
L =
  - (8\pi G)^{-1}(\Upsilon g^{00}-\Xi g^{ii})\sqrt{-g}
    + L_{GR}(g_{\mu\nu}) + L_{matter}(g_{\mu\nu},\varphi^m).
\]

\subsection{Discussion of the Derivation}

This extremely simple derivation of exact general-relativistic
symmetry in the context of a classical condensed matter theory looks
very surprising.  It may be suspected that too much is hidden behind
the innocently looking relation between Lagrange formalism and the
condensed matter equations, or somewhere in our ``weak covariant
formalism''.

Therefore, let's try to understand on a more informal level what has
happened, and where we have made the physically non-trivial assumption
which leads to the very non-trivial result -- the EEP.

A non-trivial physical assumption is, indeed, the classical Euler
equation.  This equation contains non-trivial information -- that
there is only one medium, which has no interaction, especially no
momentum exchange, with other media.  This is, indeed, a natural but
very strong physical assumption.  And from this point of view the
derivation looks quite natural: What we assume is a single, universal
medium.  What we obtain with the EEP is a single, universal type of
clocks.  An universality assumption leads to an universality result.

From point of view of parameter counting, all seems to be nice too.
We have explicitly fixed four equations, and obtain an independence
from four coordinates.  All the effective matter fields are by
construction fields of a very special type -- inner steps of freedom,
material properties, of the medium.  Therefore it is also not strange
to see them closely tied to the basic steps of freedom (density,
velocity, pressure) which define the effective gravitational field.

The non-trivial character of the existence of a Lagrange formalism
seems also worth to be mentioned here.  A Lagrange formalism leads to
a symmetry property of the equation -- they should be self-adjoint.
This is the well-known symmetry of the ``action equals reaction''
principle.  It can be seen where this symmetry has been applied if we
consider the second functional derivatives.  The EEP means that the
equations for effective matter fields $\varphi^m$ do not depend on the
preferred coordinates $X^\alpha$.  This may be proven in this way:

\[ \frac{\delta}{\delta X^\mu}     \frac{\delta S}{\delta \varphi^m}
 = \frac{\delta}{\delta \varphi^m} \frac{\delta S}{\delta X^\mu}
 = \frac{\delta}{\delta \varphi^m} \mbox{[cons. laws]} = 0 \]

Thus, we have applied here the ``action equals reaction'' symmetry of
the Lagrange formalism and the property that the fundamental classical
conservation laws (related with translational symmetry by Noether's
theorem) do not depend on the material properties $\varphi^m$.

Thus, our considerations seem to indicate that the relativistic
symmetry (EEP) we obtain is explained in a reasonable way by the
physical assumptions which have been made, especially the universality
of the medium and the special character of the ``effective matter''
steps of freedom -- assumptions which are implicit parts of the Euler
equation.

\section{Application in Condensed Matter Theory}

The connection we have established between condensed matter theory and
general relativity may be used in two directions.  We can apply
condensed matter theory to model gravity -- essentially a revival of
the old ether idea.  And we can use our understanding of general
relativity and the powerful differential-geometric machinery which has
been developed in this context in condensed matter theory.

Let's consider first the application to condensed matter theory.  If
we find a description of a condensed matter theory which fits into our
scheme, our derivation gives an important symmetry of the theory -- a
symmetry which is important for the excitations of the matter itself,
but hidden for outside observers.

The continuity equation and the Euler equation we already have in the
theory. What we need is therefore an appropriate Lagrange formulation.
Unfortunately, it is a hard problem to find Lagrangians for condensed
matter theories.  The standard Lagrangians (see, for example, the
review of Wagner \cite{Wagner}) use Lagrange multipliers to obtain the
continuity equation.  Now, what we have found is another general way
to obtain continuity and Euler equations as Euler-Lagrange equations.
What we need now are special Lagrangians for condensed matter theories
which follow this general scheme.  ``Reverse engineering'' seems to be
a reasonable way: we would have to start with a general-relativistic
matter Lagrangian, introduce the condensed matter variables, and try
to identify the matter fields with material properties of some type of
condensed matter.

In this context, it seems worth to mention that the additional terms
of the Lagrangian have a simple and natural form in the original
condensed matter variables:

\[
L = L_{GR} + L_{matter}
  - (8\pi G)^{-1}(\Upsilon \rho-\Xi (\rho v^2 + p^{ii}))
\]

The Hilbert-Einstein Lagrangian R may be interpreted as related with
inner stress.  Indeed, if there is no curvature, there exists a
coordinate transformation so that the metric becomes the Minkowski
metric, which plays the role of a stress-free reference state.  The
use of the 3D Euclidean Hilbert-Einstein Lagrangian for the
description of stress in the presence of dislocations has been
proposed by Malyshev \cite{Malyshev}.  Our approach suggests that the
other, time-dependent parts of the full Hilbert-Einstein Lagrangian
have to be taken into account too.

A solvable problem seems to be that the pressure has the wrong sign:
it should be negative definite, but usual pressure, as defined in
microscopic theories, is positive definite.  But in the classical
Euler equation pressure is defined only modulo a constant, therefore
our pressure should not be identical with microscopic pressure.

An interesting point is the symmetry group of the theories in our
class.  Looking at the basic conservation laws, which are classical,
pre-relativistic equations, it seems natural to assume Galilean
symmetry.  Instead, the symmetry group is the Poincare group.  This
Poincare symmetry is defined by the Minkowski metric
$\gamma_{\alpha\beta}$.  Thus, if we try reverse engineering, we
obtain not classical but special-relativistic condensed matter
equations.  Classical Galilean invariance we obtain only in a limit
$\Upsilon/\Xi\to\infty$.

\section{Application in Fundamental Physics: Revival of the Ether Concept}

The other direction would be to try to use a classical condensed
matter theory which fits into our scheme as the fundamental ``theory
of everything''.  Our general scheme defines a metric theory of
gravity in competition with general relativity, while the special
material properties and material equations of the model define the
matter fields and matter equations.

We obtain the following equations:

\[G^\mu_\nu  = 8\pi G (T_m)^\mu_\nu
   + (\Lambda +\gamma_{\kappa\lambda}g^{\kappa\lambda}) \delta^\mu_\nu
   - 2g^{\mu\kappa}\gamma_{\kappa\nu}\]

as well as the harmonic equations

\[ \partial_\mu  (g^{\mu\alpha}\sqrt{-g}) = 0. \]

There is also another form of the energy-momentum tensor -- the basic
equation may be simply considered as a decomposition of the full
energy-momentum tensor $g^{\mu\kappa}\sqrt{-g}$ into a part
$(T_m)^\mu_\nu$ which depends on matter fields and a part
$(T_g)^\mu_\nu$ which depends on the gravitational field:

\[(T_g)^\mu_\nu = (8\pi G)^{-1}\left(\delta^\mu_\nu(\Lambda
              	+ \gamma_{\kappa\lambda}g^{\kappa\lambda})
      	 	- G^\mu_\nu\right)\sqrt{-g}\]

Thus, the additional non-covariant terms solve the problem of general
relativity with local energy and momentum density for the
gravitational field.  We have even two equivalent forms for the
conservation laws, one with a subdivision into gravitational and
matter part, and another where the full tensor depends only on the
gravitational field.

The consideration of the predictions of the theory which differ from
general relativity is outside the aim of this paper and done elsewhere
\cite{Schmelzer}\footnote{They include a ``big bounce'' instead of a
big bang, a homogeneous dark matter term of type $p=-1/3\varepsilon$,
and stable ``frozen stars'' instead of black holes}.  But from the
previous results already follows that the Einstein equivalence
principle (which includes local Lorentz invariance, local position
invariance, and the weak equivalence principle) holds, and that in the
limit $\Xi,\Upsilon\to 0$ we obtain the Einstein equations.  The
overwhelming experimental evidence in favour of the EEP and the
Einstein equations (Solar system tests, binary pulsars) as described
by Will \cite{Will} can give only upper bounds for the parameters
$\Xi, \Upsilon$.  Therefore the theory is not in contradiction with
observation.

For most physicists it seems more objectionable that this theory is
essentially a revival of the old pre-relativistic ether theory.  This
seems contrary to the overwhelming success of relativity during this
century.  But most of this progress -- especially relativistic quantum
field theory up to the standard model -- in no way depends on the
question if there is a hidden preferred frame or not.  And, as our
theory shows, the progress in the domain of relativistic gravity can
be as well covered by an ether theory too.  So, this argument does not
seem to be decisive.

The theory we have presented here removes some old arguments against
ether theory:

\begin{itemize}

\item There was no viable theory of gravity --- we have found now a
theory of gravity with GR limit which seems viable;

\item The assumptions about the Lorentz ether have been ad hoc.  There
was no explanation of relativistic symmetry, the relativistic terms
have had ad hoc character --- the assumptions we make for our medium
seem quite natural, and we derive relativistic symmetry;

\item There was no explanation of the general character of
relativistic symmetry, the theory was only electro-magnetic --- in the
new concept the ``ether'' is universal, all matter fields describe
properties of the ether, which explains the universality of the
EEP and the gravitational field;

\item There was a violation of the ``action equals reaction''
principle: there was influence of the ether on matter, but no reverse
influence of matter on the stationary and incompressible ether --- we
have now a Lagrange formalism which guarantees the ``action equals
reaction'' principle and have a compressible, instationary medium;

\end{itemize}

But not only these old problems of ether theory, which have justified
the rejection of the ether concept, have been solved.  It is also easy
to find advantages of the ether concept -- moreover, these advantages
can be found in very different domains.  We have already mentioned
local conservation laws for energy and momentum.  There are other
domains: compatibility with quantum principles, and compatibility with
classical realism and hidden variable theories:

First, some conceptual quantum gravity problems disappear, especially
the notorious ``problem of time''.  This is a well-known fact:
``... in quantum gravity, one response to the problem of time is to
`blame' it on general relativity's allowing arbitrary foliations of
spacetime; and then to postulate a preferred frame of spacetime with
respect to which quantum theory should be written.''
\cite{Butterfield}.
\footnote{ This way to solve the problem is rejected not for physical
reasons, but because of deliberate metaphysical preference for the
standard general-relativistic spacetime interpretation: ``most general
relativists feel this response is too radical to countenance: they
regard foliation-independence as an undeniable insight of
relativity.''  \cite{Butterfield}.  Of course, these feelings are easy
to understand. At ``the root of most of the conceptual problems of
quantum gravity'' is the idea that ``a theory of quantum gravity must
have something to say about the quantum nature of space and time''
\cite{Butterfield}.  The introduction of a Newtonian background
``solves'' them in a very trivial, uninteresting way.  It does not
tell anything about quantum nature of space and time, because space
and time do not have any quantum nature in this theory -- they have
the same classical nature of a ``stage'' as in non-relativistic
Schr\"odinger theory.  The hopes to find something new, very
interesting and fundamental about space and time would be dashed.  But
nature is not obliged to fulfil such hopes.}

Another argument is related with the violation of Bell's inequality
\cite{Bell}.  It is widely accepted that experiments like Aspect's
\cite{Aspect} show the violation of Bell's inequality and, therefore,
falsify Einstein-local realistic hidden variable theories.  Usually
this is interpreted as a decisive argument against hidden variable
theories and the EPR criterion of reality.  But it can as well turned
into an argument against Einstein locality.  Indeed, if we take
classical realism (the EPR criterion \cite{EPR}) as an axiom, the
violation of Bell's inequality simply proves the existence of
superluminal causal influences.  Such influences are compatible with a
theory with preferred frame and classical causality, but not with
Einstein causality.
\footnote{
This has been mentioned by Bell \cite{Bell1}: ``the cheapest
resolution is something like going back to relativity as it was before
Einstein, when people like Lorentz and Poincare thought that there was
an aether --- a preferred frame of reference --- but that our
measuring instruments were distorted by motion in such a way that we
could no detect motion through the aether. Now, in that way you can
imagine that there is a preferred frame of reference, and in this
preferred frame of reference things go faster than light.''}

We know today that an EPR-realistic, even deterministic hidden
variable theory exists -- it is Bohmian mechanics \cite{Bohm}.  It may
be generalized into the special-relativistic domain.  In agreement
with the general properties of EPR-realistic theories, this requires a
preferred frame.  This property of Bohmian mechanics explains why it
has been widely ignored.  Now, an ether theory of gravity which gives
a preferred frame allows to extend the concept of Bohmian mechanics
into the domain of relativistic gravity.
\footnote{Note also that the existence of Bohmian mechanics proves
that QM in itself is compatible with the EPR criterion of reality, so
quantum theory gives no additional evidence again the EPR criterion.
Thus, the only argument against the EPR criterion is its contradiction
with Einstein causality.}

These considerations suggest that the old ether concept is not as
unreasonable as it seems at a first look and is worth to be considered
in more detail.
\footnote{ Note that in all the problems we have considered here
(local energy density, problem of time, EPR criterion, Bohmian
mechanics) the majority of scientists has made a different choice.  In
all cases this has been at least partially justified with the
contradiction with relativistic philosophy.  These have been
reasonable decisions in a situation where the ether concept was as
dead as possible for a scientific theory and often compared with
phlogiston theory.  But we discuss here a possible revival of this
alternative.  Whatever we present as an argument in favour of a
preferred frame we present {\em because\/} it is in contradiction with
the relativistic paradigm.  If the concept has been rejected mainly
{\em because\/} it is in contradiction with the relativistic paradigm,
this rejection cannot be taken into account in the context of this
discussion.  It would be circular reasoning. 

Thus, to evaluate above arguments in favour of a preferred frame we
cannot accept the ``majority opinion'' as it is.  Instead, we have to
reevaluate the arguments to find out if there are reasons for the
rejection of these concepts which do not depend on their contradiction
with the relativistic paradigm.  This endeavour is beyond the scope of
this article.}

\section{Conclusion}

We have found a remarkable connection between a simple class of
classical condensed matter theories and a metric theory of gravity
which fulfils the EEP and gives in some limit the Einstein equations
of general relativity.  It is a simple and beautiful relation which
combines in a natural way two very beautiful coordinate-dependent
elements of general relativity -- harmonic coordinates and the ADM
decomposition -- with the classical equations of condensed matter
theory.

This relation suggests applications in two directions.  It may be used
to apply the apparatus of general relativity in condensed matter
theory.  We can also use classical condensed matter theory as a model
for fundamental physics below Planck scale.  This may be a use as a
toy model, similar to the use of acoustic ``dumb holes'' as models for
the study of Hawking radiation \cite{Rosu}.  But it seems not
unreasonable to think also about a complete revival of
pre-relativistic ether concepts.  We have found that many old problems
of the Lorentz ether are solved by our new approach, and we have also
noted several independent arguments in favour of a preferred frame:
local energy-momentum densities for gravity, the problem of time in
quantum gravity, the EPR criterion of reality and Bohmian mechanics.

Future has to show which of these possibilities will be successful.
Anyway, the relation between condensed matter theory and the theory of
gravity we have found here seems too beautiful to be without any
physical importance.

\begin{appendix}

\section{Can every Lagrangian be rewritten in weak covariant form?}
\label{appGeneral}

There has been some confusion about the role of covariance in general
relativity.  Initially it was thought by Einstein that general
covariance is a special property of general relativity.  Later, it has
been observed that other physical theories allow a covariant
description too.  The classical way to do this for special relativity
(see \cite{Fock}) is to introduce the background metric
$\eta_{\mu\nu}(x)$ as an independent field and to describe it by the
covariant equation $R^\mu_{\nu\kappa\lambda}[\eta]=0$.  For Newton's
theory of gravity a covariant description can be found in \cite{MTW},
\S 12.4.  The general thesis that every physical theory may be
reformulated in a general covariant way was proposed by Kretschmann
\cite{Kretschmann}.

In our derivation we have to assume that the Lagrangian is given in
our ``weak covariant'' way.  Now, the question we want to discuss here
is if this is a non-trivial restriction or not.  Thus, we have here a
similar but more specific problem: We have to consider condensed
matter theories with Lagrange formalism, and have to rewrite them in a
covariant way with very special variables -- the preferred coordinates
$X^\alpha$ -- so that every dependence on the Newtonian background is
described by these variables $X^\alpha$.  Our hypothesis is that this
is possible:

\begin{center}\em
The Lagrangian of every physical reasonable condensed matter theory
with Lagrange formalism may be rewritten in ``weak covariant''
form.
\end{center}

Unfortunately we don't know what may appear in the ``most general
reasonable theory'' as a variable.  Therefore, in this general form
this thesis seems unprovable in principle.  All we can do is to
present ways to transform some more or less general classes of
Lagrangians into weak covariant form.

Now, for the classical geometric objects related with the Newtonian
framework it is easy to find a weak covariant formulation:

\begin{itemize}

\item The preferred foliation is immediately defined by the coordinate
$T(x)$.

\item The Euclidean background metric: a scalar product
$(u,v)=\delta_{ij}u^iv^j$ may be presented in weak covariant form as
$(u,v)=\delta_{ij}X^i_{,\mu}X^j_{,\nu}u^\mu v^\nu$.

\item Absolute time metric: Similarly, the (degenerate) metric of
distance in time may be defined by $(u,v)=T_{,\mu}T_{,\nu}u^\mu
v^\nu$.

\item The preferred coordinates also define tetrad and cotetrad
fields: $dX^\alpha = X^\alpha_{,\mu} dx^\mu$, for the vector fields
$\partial/\partial X^\alpha =
(J^{-1})^\mu_\alpha(X^\beta_{,\nu})\partial/\partial x^\mu$ we need
the inverse Jacobi matrix $J^{-1}$, which is a rational function of
the $X^\beta_{,\nu}$.

\item Arbitrary upper tensor components $t^{\alpha_1,..\alpha_n}$ may
be transformed into $X^{\alpha_1}_{,\mu_1}\cdots
X^{\alpha_n}_{,{\mu_n}}t^{\mu_1,..\mu_n}$.  For lower indices we have
to use again the inverse Jacobi matrix $J^{-1}$.

\item Together with the background metrics we can define also the
related covariant derivatives.

\end{itemize}

Now, these examples seem to be sufficient to justify our hypothesis.
Of course, if we want to transform a given non-covariant function into
a weak covariant form, we have no general algorithm.  For example, a
scalar in space in a non-relativistic theory may be a spacetime
scalar, the time component of a four-vector, and so on for tensors.
But this is not the problem we have to solve here.  For our hypothesis
it is sufficient that there exists at least one way.

Last not least, let's note that our hypothesis is not essential for
our derivation.  Simply, if the hypothesis is false, our assumption
that the Lagrangian is given in weak covariant form is really
non-trivial.  This would be some loss of generality and therefore of
beauty of the derivation, but in no way fatal for the main results of
the paper.  The considerations given here already show that the class
of theories which allows a weak covariant formulation is quite large.

\section{Remarks about the special geometric nature of coordinates}
\label{appCoordinates}

In our formalism we handle the preferred coordinates like usual
fields.  Of course, every valid set of coordinates $X^\alpha(x)$
defines a valid field configuration.  But the reverse is not true.  To
define a valid set of coordinates, the functions $X^\alpha(x)$ have to
fulfil special local and global restrictions: the Jacobi matrix should
be non-degenerated everywhere, and they should fulfil special boundary
conditions.

Thus, we have to be careful and to check how this special geometric
nature of the preferred coordinates influences various questions.

\subsection{Justification of the Euler-Lagrange formalism}

The description of the Lagrange function in weak covariant form is, in
comparison with its non-covariant description, only another way to
describe the same minimum problem in other variables.  So, to solve
this minimum problem we can try to apply the standard variational
calculus.

Now, there is a subtle point which has to be addressed.  The point is
that not all variations $\delta X^\alpha(x)$ are allowed -- only the
subset with the property that $X^\alpha(x)+\delta X^\alpha(x)$ defines
valid global coordinates.  Therefore, to justify the application of
the standard Euler-Lagrange formalism we have to check if this subset
of allowed variations is large enough to give the classical
Euler-Lagrange equations.

Fortunately this is the case.  To obtain the Euler-Lagrange equations
we need only variations with compact support, so we don't have a
problem with the different boundary conditions.  Moreover, for
sufficiently smooth variations ($\delta X^\alpha \in C^1({\mathbb R}^4)$
is sufficient) there is an $\varepsilon$ so that
$X^\alpha(x)+\varepsilon\delta X^\alpha(x)$ remains to be a system of
coordinates: indeed, once we have an upper bound for the derivatives
of $\delta X^\alpha$, we can make the derivatives of
$\varepsilon\delta X^\alpha$ arbitrary small, especially small enough
to leave the Jacobi matrix of $X^\alpha+\varepsilon\delta X^\alpha$
non-degenerated.

Now, to obtain the Euler-Lagrange equations we need only small
variations, so, the geometrical restrictions on global coordinates do
not influence the derivation of the Euler-Lagrange equations for the
$X^\alpha(x)$.  So, for the preferred coordinates we obtain usual
Euler-Lagrange equation, as if they were usual fields, despite their
special geometric nature.

\subsection{Comparison with ``GR with clock fields''}

This does not mean that the special geometric nature of the preferred
coordinates is unimportant.  Instead, this special nature of the
fields $X^\alpha(x)$ is a very important part of the definition of our
condensed matter theories.

In this context it is useful to compare our condensed matter theories
with the theory we obtain by forgetting this special geometric nature.
We obtain a general-relativistic theory with four scalar fields
$U^\alpha(x)$ and Lagrangian

\[L = -(8\pi G)^{-1}\gamma_{\alpha\beta}U^\alpha_{,\mu}U^\beta_{,\nu}
       	g^{\mu\nu}\sqrt{-g}
    + L_{GR}(g_{\mu\nu}) + L_{matter}(g_{\mu\nu},\varphi^m).
\]

A similar theory with ``clock fields'' in general relativity has been
considered by Kuchar \cite{Kuchar}.  Now, we have the same Lagrangian
and the same Euler-Lagrange equations.  Nonetheless, above theories
are completely different -- they don't have even a single common
solution.  Indeed, we have completely different boundary conditions.
For a standard field theory with scalar fields, the natural boundary
condition is that the solution remains bounded.  Instead, the
coordinates should be unbounded on a complete solution.

Nonetheless, even if above theories have not even a single common
solution, the question how to distinguish above theories
by observation remains very complicate.

\end{appendix}

\end{document}